\documentstyle[aps,prl]{revtex}
\input  epsf
\begin{document}
\draft

\twocolumn[\hsize\textwidth\columnwidth\hsize\csname@twocolumnfalse%
\endcsname
\title{Configuration space in electron glasses}

\author{A. P\'erez--Garrido, M. Ortu\~no, A. M. Somoza 
         and A. D\'{\i}az--S\'anchez
}

\address{Departamento de F\'{\i}sica, Universidad de 
Murcia,
Murcia 30.071, Spain}

\maketitle

\begin{abstract}
{ We study numerically  the configuration space at low energy of electron 
glasses. We consider systems with Coulomb interactions, short--range interactions and no
interactions.
First, we calculate the integrated density of configurations as a function of energy. 
At a given
energy, this density is smaller for Coulomb glasses than for short-range systems, which
in turn is smaller than for non--interacting systems. We analyze how the site
occupancy varies with the number of configurations. Through this study we
estimate the number of particles involved in a typical low--energy transition between
configurations. This number increases with system size for long range interactions,
while it is basically constant for a short-range interaction. Finally we calculate the
density of metastable configurations, {\it i.e.} valleys, classified according to their
degree of stability. 
}
\end{abstract}

]

\section{Introduction}

The term Coulomb glass refers to Anderson insulators with Coulomb 
interactions between the localized electrons, and it has established itself
as an important model in the study of the electronics properties of
insulators \cite{PO85}. It assumes that quantum energies $t$ arising from tunneling 
are much smaller than the other important energies in the problem, 
{\it i.e.} Coulomb interactions and random energy fluctuations. 
In this situation the Hubbard energy is very large and practically
forbids double occupation of sites. 
The interesting case then corresponds to a number of electrons roughly 
half the number of sites. 

Until a few years ago, most studies of the Coulomb glass consisted 
mainly of finding the ground state of the system, and considering from 
there on only one--particle transitions \cite{BE79,DL84,SE84}. The main problem arising
with this is the neglection of many--body effects. Several
experimental \cite{MK95,Ad89,PS93} 
and numerical \cite{PO97} results give evidences that electronic correlations 
are very important in these
systems and they must be taken into account.  
More recently, methods were 
developed to obtain an almost
complete set of low--lying states of the Coulomb glass \cite{MP91,ST93}.
This process allows for a much more detailed consideration of many--body
effects.

A proper understanding of the low-energy configuration space is
extremely important for the study of most physical properties. 
Previous works on Coulomb glasses have never focus on the properties 
and relations between
configurations. We only have an indirect knowledge of them through the
analysis of different physical problems. In this work we perform a
systematic study of the low-energy configurations of three-dimensional
Coulomb glasses. For the sake of comparison, we
have also computed the same properties for non-interacting systems and
for disordered systems with short range interactions.

In the next section, we introduce the model used and details of the
numerical simulations. In section III, we study the entropy of the
systems and compare to that obtained by two different analytical models.
In section IV, 
we obtain the
average difference in occupancy between the ground state and configurations as a
function of energy, and in section V 
 we calculate the average  number of electrons
participating in a typical low--energy transition. In section VI we study the density of
valleys in configuration space. Finally, in section VII we extract some conclusions.

\section{Numerical model}
We consider three--dimensional systems deep in the insulating regime,
where the relevant energies in the problem are the Coulomb interaction
and the disorder energy. This
corresponds to the condition $a \ll r_0$, where $a$ is the localization 
radius, $r_0=(4\pi n/3)^{-1/3}$ is the
typical distance between sites and $n$ the concentration of sites.

We use the standard tight--binding Hamiltonian:
\begin{equation}
H=\sum_i\epsilon_i (n_i-K) +\sum_{i<j} V_{ij}, \label{hamil}
\end{equation}
$\epsilon_i$ are the site random energies, uniformly distributed at 
random in the interval $(-W/2,W/2)$, $W$ being the disorder strength.
We  take the number of electrons to be  half the number of sites and 
$K=0.5$. The sites are arranged at random, but with a minimum separation
between them, which we choose to be $0.5r_0$. We take $e^2/r_0$ as our 
unit of energy and $r_0$ as our unit of distance.
$V_{ij}$ is the interaction potential. In this work we use three types
of potentials, coulombic, short range and non--interacting.
The Coulomb potential is given by the expression
\begin{equation}
V_{ij}=\frac{\left(n_i-K\right)\left(n_j-K\right)}{r_{ij}},
\end{equation}
where $r_{ij}$ is the separation between sites $i$ and $j$. For the short 
range potential we choose
\begin{equation}
V_{ij}=\left(n_i-K\right)\left(n_j-K\right)\left(\frac{\sigma}{r_{ij}}
\right)^4,
\end{equation}
$\sigma$ was taken equal to 0.7. Finally, the non--interacting case
simply corresponds to $V_{ij}=0$.

A numerical algorithm developed  to obtain the ground state and
the lowest energy many--particle configurations of these systems has been 
used for the calculation of low energy states \cite{OP95}.
This algorithm considers simultaneous  many--electron transitions constructed
from (of the order of ten) independent one--electron transitions and
perform the combination that relaxes more energy.
Starting from a random initial configuration, we repeat the relaxation
procedure until we cannot reduce the energy any more. The whole scheme
is repeated for different initial random
configurations of the charges until the configuration of lowest energy is
found ten times. The configurations thus generated were memorized in terms
of site occupation numbers (0 or 1) and of energy, whenever this was less
than the highest energy configuration in memory storage.
We complete the set of low--energy configurations by generating all
the states that differ by one-- or two--electron transitions from any
configuration stored.

\section{Number of Configurations}

The first magnitude to study is the number of configurations as a
function of energy. 
Let us call density of states to the number of one--particle states per unit
energy, and density of configurations $n(E)$ to the number of 
many--particle configurations per unit energy. 
In particular, we consider the integrated density of configurations $N(E)$, {\it i.e.}
 the number of 
configurations  with energy smaller than or equal to $E$. 
If the occupation probability of all configurations up to an energy $E$ 
is the same, the entropy, $S$, is then equal to $k_{\rm B}\ln N(E)$,
where $k_{\rm B}$ is the Boltzmann constant which we take equal to 1. 
The number of configurations strongly depends on the size of the sample,
but this dependence can be roughly taken into account by an adequate
normalization of the energy. The idea is to measure the energy in terms
of the average energy spacing between single particle states near the
Fermi level (in the absence of interactions):
\begin{equation}
E={\mathcal E} V/W,
\label{En}
\end{equation}
where $E$ is the normalized energy, $V$ the volume of the system, $W$ the disorder
strength and $\mathcal{E}$ the energy to be normalized.  This procedure provides a
density of configurations
independent of dimensionality and system size for  non interacting systems. 

\begin{figure}
\epsfxsize=\hsize
\begin{center}
\leavevmode
\epsfbox{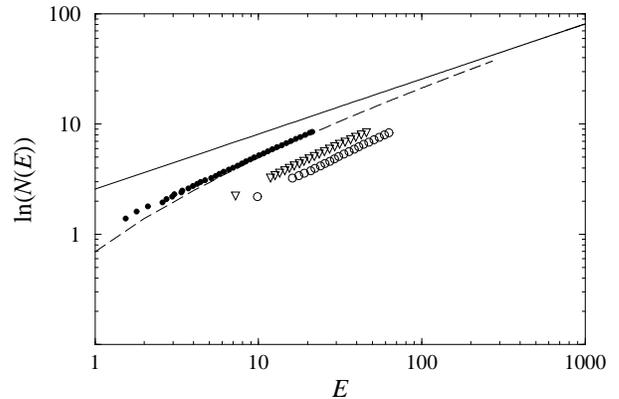}
\end{center}
\caption{$\ln N(E)$ versus energy for systems with no interactions
(solid circles), short range interactions (triangles) and Coulomb 
interactions (empty circles). The solid line corresponds to the 
standard result, Eq.\ (\ref{entropy}), and the dashed line to our discrete
model, Eq.\ (\ref{integrated}).}
\end{figure}

In Fig.\ 1 we plot on a double logarithmic scale the average of the 
entropy as a function of the normalized energy, {\it i.e. } measured in terms
of the one-particle energy spacing. Averages are taken over twenty 
samples. The solid dots correspond to the
non-interacting case, the triangles to a system with short-range 
interactions, and the empty dots to a Coulomb glass. 
As we have previously mentioned, the results for the non-interacting 
case are exactly independent of the sample size. The two interacting
cases correspond to averages over samples with a number of sites equal 
to $N_{\rm s}=900$. Although, in principle, the results for interacting
systems depend on size, the variation is very small and we have 
preferred not to show them for several sizes for the sake of clarity. 
In all cases considered, the fluctuations from sample to sample are
large, even in the logarithmic scale considered, but averages are
relatively well behaved.

For non-interacting systems it is possible to make theoretical calculations
for the entropy. The classical result obtained from statistical
mechanics for an infinite system with a constant density of states
corresponds to the solid line in Fig.\ 1. The entropy is proportional to 
$\sqrt{E}$, and so this is a straight line with slope equal to 0.5. 
Although the previous result is  well known, 
we deduce it below in order to express it in our notation and
because the method of calculation can be extended to other non-standard
magnitudes as we will see later on. The dashed line connects the points obtained
from a finite discrete model that we also describe in the next subsection.
All systems have a integrated density of the form  
$N(E)\propto \exp (E/E_0)^\alpha$. In all cases, for the sizes 
considered we checked that $\alpha\approx 0.7$.
The value of the exponent, $\alpha$, is deeply  affected by
finite size effect. 
For interacting systems we
are not sure of the value the exponent  in the thermodynamic limit.
The similarity between the interacting and non--interacting
systems for the energy ranges considered suggests that $\alpha=0.5$ might also 
correspond to the thermodynamic limit for the interacting cases.

\subsection{Theoretical models}

We want to derive the entropy for a system of
non interacting electrons with a constant density of states.
It is possible to obtain analytically the entropy and other properties  
of this system for two models. Firstly,
we study a continuous model using one--particle theory. Secondly, we
investigate a discrete model of equally spaced levels by counting 
directly the number of configurations.

We first assume a continuous model with a constant density of 
states for the whole range of energies of interest, $g(\epsilon)=g_0$. 
The chemical potential $\mu$ and the Fermi level $\epsilon_f$ are equal 
to zero by symmetry. 
At a given temperature $T$, the average occupation number of a state 
with energy $\epsilon$ is:
\begin{equation}
n_\epsilon=\frac{1}{1+\exp( \epsilon /T)}.
\end{equation}
So the mean energy  $\langle E \rangle$ of the system, measured from the ground state
energy, is
\begin{equation}
\langle E \rangle = \int _{-\infty}^{\infty}
(n_\epsilon-\theta(\epsilon))Vg_0\epsilon\,d\epsilon =\frac{\pi^2Vg_0T^2}{6},
\label{energia}
\end{equation}
where $\theta(\epsilon)$ is the step function, which takes into account the
ground state energy, and $V$ is the volume of the system. Eq.\
(\ref{energia}) allows us to change between temperature and mean energy.

The entropy $S$ of a non-interacting system can be calculated from the 
average site occupation through the expression \cite{H63}
\begin{equation}
S=\int_{-\infty}^\infty Vg_0 \left\{ 
n_\epsilon \ln\left( \frac{1}{n_\epsilon}-1 \right) -
\ln \left( 1- {n_\epsilon} \right)
\right\} d\epsilon .
\label{integral}
\end{equation}
Integrating this expression and using Eq.\ (\ref{energia}), one obtains
the entropy as a function of energy:
\begin{equation}
S=\sqrt{\frac{2\pi^2Vg_0\langle E \rangle}{3}}.
\label{entropy}
\end{equation}

Eq.\ (\ref{entropy}) is plotted as the solid line of Fig.\ 1. 
We can see that the numerical results do not match the theoretical one
for the continuous case, Eq.\ (\ref{entropy}). To check the importance of the discrete
nature of our numerical systems we develop a discrete model.  
We now describe this discrete model for the number of configurations
of non-interacting electrons with equally spaced one-particle energy
levels. The dashed line in Fig.\ 1 represents $\log N(E)$ for this model. 

Each one--particle state is labeled according to its position in the energy
scale. The highest occupied level in the ground state is assigned to the
zero of our scale. In the ground state, occupied levels correspond to 
negative numbers, while empty levels to positive numbers. A state
$k$ has an energy equal to $k\Delta\epsilon$, where $\Delta\epsilon$ is the 
constant energy spacing between levels, which we take equal to 1.
The density of configurations $n(E)$ can be decomposed as
\begin{equation}
n(E)=\sum_{i=1}^\infty n_i(E),
\label{ne}
\end{equation}
where $n_i(E)$ is the number of configurations with $i$ excitations ($i$ 
electrons above Fermi level and $i$ holes below it) 
and with a total excitation energy equal to $E$. 
The excitation energy of a transition from level $-m$ (with $m\ge 0$), 
to level $l$ ($l>0$), is equal to $l-m$, recall that $\Delta\epsilon=1$.
For one electron jumps, $n_1(k)$ is simply
\begin{equation}
n_1(k)=k,
\end{equation}
since we have $k$ different possibles jumps of length $k$.

For $i=2$ (two excitations) the possibles jumps can be  from $\epsilon_{-n}$ to
$\epsilon_j$ and from $\epsilon_{-p}$ to $\epsilon_l$, with the conditions
$n\ne p$, $j\ne l$ (from two different states to two other different states),
$n,p\ge 0$ and $j,l>0$. The total energy is the sum of individual energy
\begin{equation}
k=(j-n)+(l-p).
\label{k}
\end{equation}
We rewrite equation (\ref{k}) as
\begin{equation}
k=j+l-(n+p).
\label{k2}
\end{equation}
$n_2(k)$ is, then, the number of possible combinations of $j,l,n$ and
$p$ verifying expression (\ref{k2}). Using the function $Q_i$ explained
below and (\ref{k2}) we arrive at
\begin{equation}
n_2(k)=\sum_{j=3}^{k-1}Q_2(j)Q_2(k-j+2).
\label{n2}
\end{equation}

For an arbitrary $i$, the number $n_i(k)$ of $i$ transitions with 
a total energy $k$ becomes
\begin{equation}
n_i(k)=\sum_{j=i(i+1)/2}^{k-i(i-1)/2}Q_i(j)Q_i(k-j+i).
\label{ni}
\end{equation}

The integrated density of configurations $N(k)$ is
\begin{equation}
N(k)=1+\sum_{i=1}^kn(i),
\label{integrated}
\end{equation}
where $n(i)$ is given by Eq.\ (\ref{ne}) and  where we
have added 1 to  count the ground state.

We have to
obtain the number $Q_i(m)$ of possible combinations of $i$ integers
whose sum is equal to $m$, and subject to the constraint that they must 
all be different from each other and from zero. 
When $i=1$ we have the trivial
case
\begin{equation}
Q_1(m)=1, 
\end{equation}
independently of $m$. It is easy to check that for $i=2$ that function becomes
\begin{equation}
Q_2(m)={\rm Int}\left( \frac{m-1}{2}\right),
\end{equation}
where ${\rm Int}(m)$ truncates $m$ to  its integer part. When \hbox{$i>2$}, the
function $Q_i(m)$ is defined in a recursive form as
\begin{equation}
Q_i(m)=\sum_{j=1}^{{\rm Int}(m/i)-1} Q_{i-1}(m-ij).
\end{equation}

A line connecting the points given by 
the integrated density of configurations $N(E)$ of our discrete model,
obtained from Eqs. (\ref{integrated}), (\ref{ne}) and  (\ref{ni}) 
(where $E\equiv k\cdot\Delta \epsilon$), 
has been plotted in Fig.\ 1 
(dashed line) as a function of energy. 
We can see that our predictions with the discrete model fit very well
our numerical results for systems with no interactions (solid dots). 
Furthermore, the number of configurations of the discrete model tends to the
number of configurations of the continuous model, Eq.\ (\ref{entropy}), represented
by the solid line in Fig.\ 1.

\section{Difference in occupation}

An interesting quantity is the number of electrons that have to jump from the ground
state to reach a given configuration, {\it i.e.}, half the Hamming distance between this
configuration and the ground state. We consider the average  $\langle i \rangle$ of this
quantity over small energy intervals and disorder realizations.
For a  non-interacting system this magnitude corresponds to the
average occupation of states  above the Fermi level, which for
a (continuous) constant density of states is given by:
\begin{equation}
\langle i \rangle = \int_0^\infty\frac{g_0}{1+\exp (\epsilon/T)}
d\epsilon = g_0 T \ln 2.
\end{equation}
We use Eq.\ (\ref{energia}) to rewrite $\langle i \rangle$ as a function of the
mean energy, instead of temperature $T$,
$\langle E \rangle$,
\begin{equation}
\langle i \rangle = \frac{\ln 2}{\pi}\sqrt{6\langle E \rangle g_0}.
\label{icon}
\end{equation}
We can also calculate
this quantity using our discrete model, explained before, and we
get
\begin{equation}
\langle i \rangle =
N^{-1}(E)\left( \sum_{i=1}^k\sum_{j=1}^\infty j n_j(i)\right),
\label{idis}
\end{equation}
where $N(E)$ is given by Eq.\ (\ref{integrated}) and $n_i(E)$ by Eq.\ (\ref{ni}).
In Fig.\ 2 we plot $\langle i \rangle$ given by
 Eq.\ (\ref{icon}) (solid line) and Eq.\ (\ref{idis}) 
(dashed line)  as a function of energy. 
In the same figure we also represent the results obtained by numerical calculations, 
 for a non interacting system (solid dots) and 
for coulombic systems (triangles correspond to 465 sites and 
empty squares to 899 sites). The 
results for non-interacting systems agree
fairly well with the  discrete model and, using the normalized energy, 
are independent of size.
On the other hand, the mean number of electrons above the Fermi level 
is larger 
for systems with interactions  due to the electronic 
correlations, confirming our results of the previous section.
It is interesting to note that the results for the interacting case 
are also independent of the size of the
system when plotted, as in Fig.\ 2, using the normalized energy $E$, Eq.\ (\ref{En})
at least for the sizes considered.
Both curves for Coulomb interacting systems fit to an expression of the form
\begin{equation}
\langle i \rangle = C E^\alpha ,
\end{equation}
with an exponent $\alpha$ roughly equal to 0.25, half the value for non--interacting
systems.

\begin{figure}
\epsfxsize=\hsize
\begin{center}
\leavevmode
\epsfbox{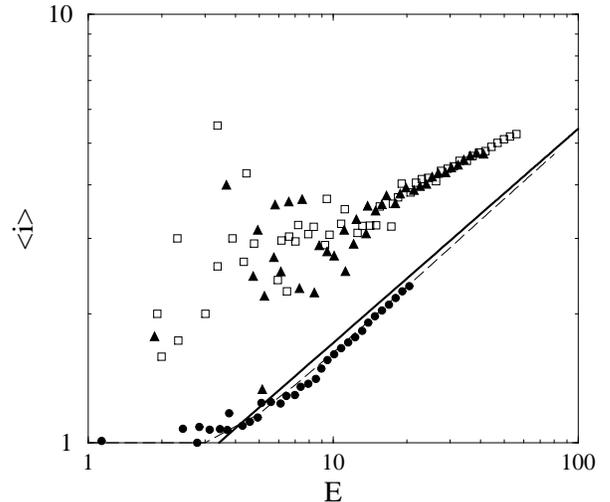}
\end{center}
\caption{Number of electron above the Fermi level as a function of energy. Solid dots
correspond to a non--interacting system. Triangles correspond to a system of 465 sites  
 and empty squares correspond to  a system of 899 sites, 
both of them with Coulomb interactions. 
The dashed line is given by the discrete model and the solid line by the continuous one.}
\end{figure}

\section{Effective number of electrons participating in low--energy transitions}

Let us consider that our system is in any of the $N(E)$ configurations
 up to energy $E$ with equal probability and let us study how the
 site occupancy depends on the interactions present.
For the non--interacting system all configuration are related to at least another
lower energy configuration by a single electron jump. This is no longer true for
interacting systems. We can  generalize this idea
through the concept of the Hamming distance between two configurations, which is defined 
as the number of sites with different occupation in both configurations.
In the non--interacting case, for any given configuration there is always at least one
lower energy configuration whose Hamming distance is equal to 2.
 We note that in the previous section we calculated the Hamming
distance to the ground state measured in the number of electron jumps needed to reach
this state from a given configuration. Now we try to estimate 
for interacting systems the (average)
 minimum Hamming
distance from a given configuration to any lower energy configuration. Obviously this quantity will strongly affect the structure
of phase space and in particular will influence the  number of sites whose occupation
changes in the $N(E)$ configurations.

For non--interacting systems following a Fermi--Dirac statistic, the importance of the
number of sites whose occupation changes is reflected in the fact that, in the
thermodynamic limit, the entropy $S=\ln N(E)$ can be expressed as a sum over  the sites:
\begin{equation}
S_N= -\sum_i^n\left(\langle s_i\rangle\ln\langle s_i\rangle +
\langle 1-s_i\rangle\ln\langle 1-s_i\rangle\right) ,
\label{S2}
\end{equation}
where $\langle s_i\rangle$ is the average occupation number at site $i$ and
$n$ is the number of sites in the sample (note that sites with either 
$\langle s_i\rangle=0$ or 1 do not contribute to $S_N$).
In general, for an interacting system, 
$S_N$ does not correspond to the real entropy.
A very simple interacting case where it is possible to obtain a similar
relation is a system 
whose elementary excitations consist of simultaneous jumps of $q$ electrons. If these
excitations are independent, the 
number of sites contributing to the 
sum in Eq.\ (\ref{S2}) is $q$ times the number for the non--interacting case. 
Thus, $S_N=q\ln N(E)$ is proportional to the real entropy.
We could propose $S_N/\ln N(E)$ as a rough quantity to estimate $q$
but it is affected by the finite size 
effects observed in Fig.\ 1. To avoid this finite size effects, we
propose the quantity $S_N/S_N^{\rm (NI)}$, where $S_N^{\rm (NI)}$ is 
$S_N$ for a non--interacting system, as a measure of
the number of electrons participating in a low--energy transition.
In Fig.\ 3, we plot $S_N/S_N^{\rm (NI)}$ as a function of $\ln (N(E))$ 
 for a system with short--range 
interactions of 899 sites 
(solids triangles), and for three sizes of a coulombic system: 899 sites (empty dots),
465 sites (diamonds) and 248 sites (empty squares).
$S_N$ depends on the size of the system when interactions are present. For systems
with short range interactions this dependence is much weaker and we only  
plot one size in Fig.\ 3 for the sake of clarity. 
These results nicely show the importance of correlations in the low--energy
configurations of interacting systems.

\begin{figure}
\epsfxsize=\hsize
\begin{center}
\leavevmode
\epsfbox{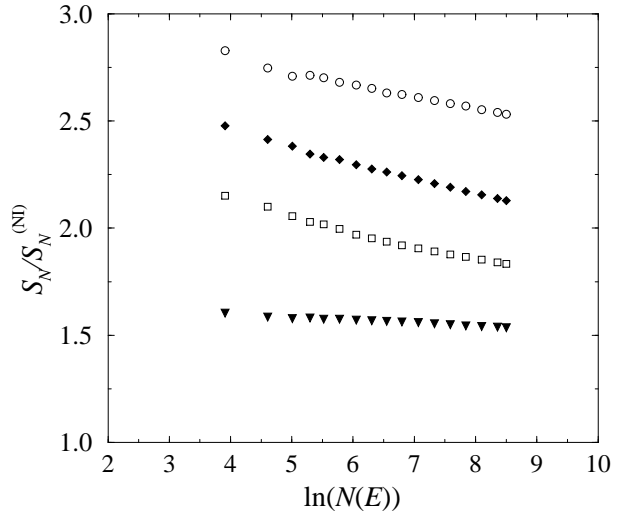}
\end{center}
\caption{$S_N/S_N^{\rm (NI)}$ as a function of $\ln (N(E))$, where $N(E)$ 
is the number of low--energy
configurations considered for systems with 
short range interactions of 899 sites (solids triangles) and systems with
 Coulomb interactions of 899 sites (empty dots), 465 sites (diamonds)
and 248 sites (empty squares).}
\label{S}
\end{figure}


\section{Valleys in configuration space}

To finalize, we study some aspects of the valley structure of the space of low--energy
configurations for systems with Coulomb interactions. 
In this space we can define a separation (which does not have  the
metric properties of a distance) between configurations $I$ and $J$, with energies
$E_I$ and $E_J$, respectively, through their transition rate $\omega_{IJ}$, taken to be
equal to \cite{PO85}: 
\begin{equation}
\omega_{IJ}=\frac{1}{\tau_0}\exp \left\{-2\sum r_{ij}/a\right\}
\exp\left\{
\frac{E_J-E_I}{kT}
\right\},
\label{w}
\end{equation}
for $E_J>E_I$, and without the second exponential for the opposite case. In Eq.
(\ref{w}),
$\tau_0$ is the inverse phonon frequency, of the order of $10^{-13}$ s, $a$ is the
localization radius, and $\sum r_{ij}$ is the minimized sum of the hopping lengths of
the electrons participating in the transition.

\begin{figure}
\epsfxsize=0.65\hsize
\begin{center}
\leavevmode
\epsfbox{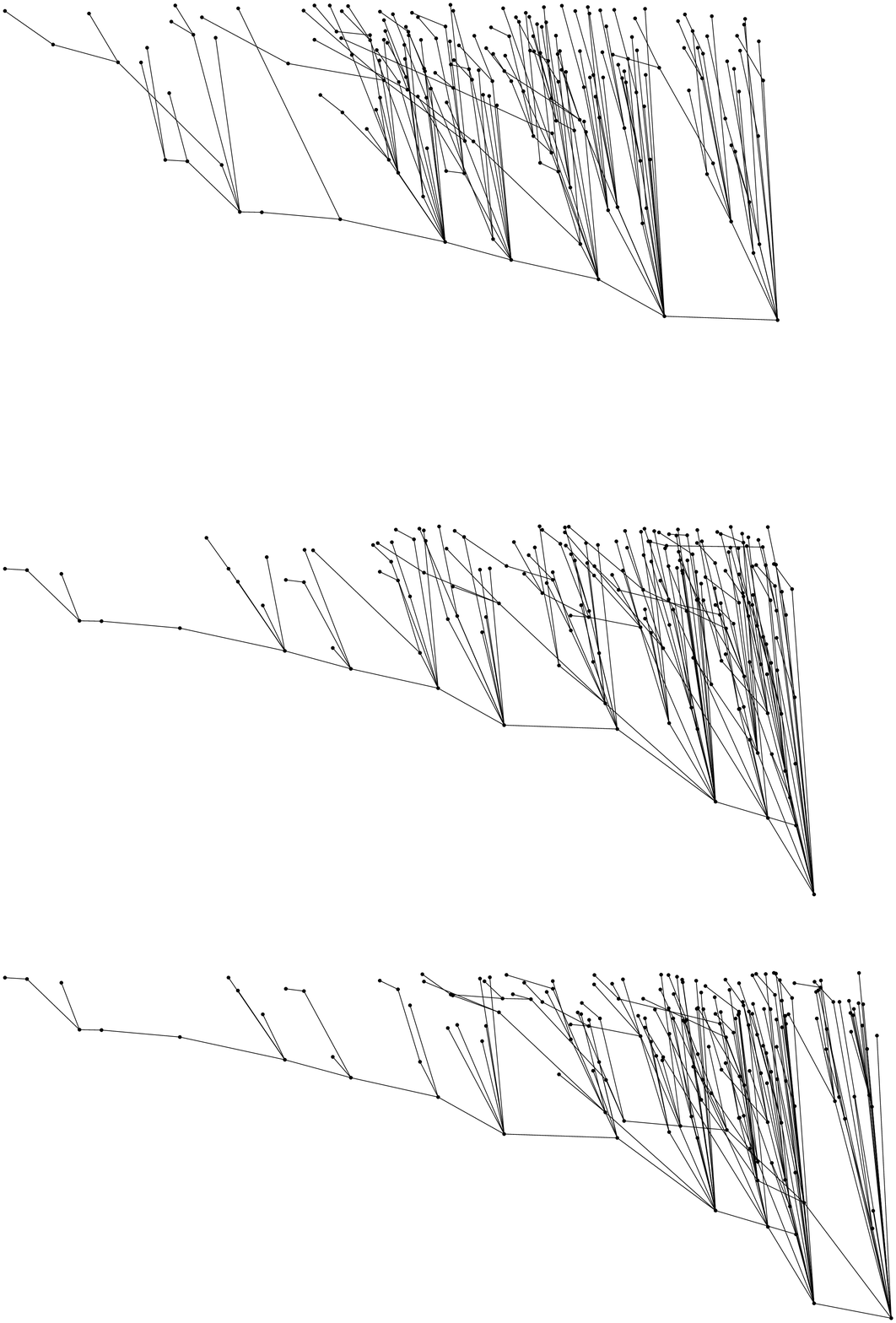}
\end{center}
\caption{Relaxation-trees in configuration  space. Each dot represents
 a configuration and each line the fastest transition from the upper configuration,
vertical distances represent their energies, being
their mutual horizontal distance the logarithm of the 
transition time. 
The bottom of the valley is a different configuration for each case
represented. The configurations considered as bottom of the valley are
 the ground state (figure at the bottom), 
the second configuration and the eighteenth configuration (figure
at the top). 
}
\end{figure}

A way to study the valley structure is through the relaxation processes. More
specifically, we assume a very small temperature so that from a given excited 
configuration we always go downwards in energy 
and we consider only the fastest transition rate at each step. Thus, we associate each
configuration with a lower energy one and we construct in this way relaxation--trees.
In Fig.\ 4 we show three different relaxation--trees in configuration space for
the same sample.
All transition jumps considered are plotted as lines where the horizontal distance is
proportional to the logarithm of the transition time and the vertical one to the jump in
energy.
The bottom valley of the three cases represented are (in ascending order)
 the ground state, the second and the eighteenth configuration. 
 At a qualitative level at least we can appreciate certain self-similarity
between the trees. 

\begin{figure}
\epsfxsize=\hsize
\begin{center}
\leavevmode
\epsfbox{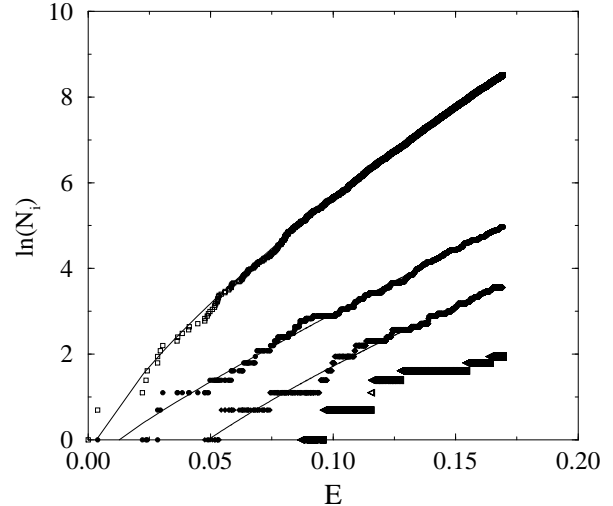}
\end{center}
\caption{Integrated densities of configurations $N_i(E)$ versus energy $E$. 
Empty squares corresponds to the
integrated density of configurations, $i=0$, while the other curves correspond to 
integrated densities
of metastable configurations with different characteristic number of electrons: 2 (solid
dots), 3 (solid diamonds) and 4 (empty triangles).
All curves fit fairly well the expression $N(E)=C\exp(E/E_0)^\alpha$, with different
$C$ and $E_0$ but
with an exponent $\alpha$  close to $0.7$ in all cases. }
\label{fi}
\end{figure}

Each metastable configuration,
whose fastest transition 
downward in energy
is carried out by the simultaneous jump of two or more electrons,
corresponds to 
the bottom of a valley.
It can be characterized by the number of electrons, $i$, participating in the fastest
transition downward in energy. Its corresponding valley is the set of configurations
that relax to it. For each characteristic number, $i$,
let us  define the integrated density of valleys, $N_i(E)$, 
as  the number of $i$--valleys  up
to an energy $E$. The standard integrated density of configurations is, obviously,
$N_0(E)=\sum_{i=1}N_i(E)$, where in order to use a closed notation we denote by $N_0$ to
the standard integrated density of configurations 
and by $N_1$ to the integrated density of configurations
whose fastest transition is a one electron jump.
We have numerically checked that these integrated densities fit fairly well an
expression of the form
\begin{equation}
N_i(E) = C\exp \left\{ (E/E_0)^\alpha \right\},
\label{NE}
\end{equation}
where $C$ and $E_0$ depend on the characteristic number of electrons. 
On the other hand,
the exponent $\alpha$ does not seem to depend on the characteristic number of electrons 
although it presents very large fluctuations from sample to sample.  
For the range of energies studied this exponent is roughly 
 0.7, the exponent for the integrated density of configurations $N_0(E)$
at the energy interval considered. 
We know from a previous section that $\alpha=0.5$ in the thermodynamic
limit.
The proportion of metastable states with respect to normal configurations 
decreases with increasing energy. 
In Fig.\ \ref{fi}, the density of configurations $N_0(E)$ (empty squares) 
and the density of valleys
stables against 2 (solid dots), 3 (solid diamonds) and 4 (solid triangles) 
simultaneous transitions are plotted  
as a function of energy, $E$. All curves fit fairly well the expression 
$N(E)=C\exp(E/E_0)^\alpha$, with different
$C$ and $E_0$ but
with an exponent $\alpha$ very close to $0.7$ in all cases.

\section{Conclusions}
We have  calculated the integrated density of configurations for systems with 
no interactions, short range interactions and Coulomb interactions.
A comparation of our numerical results with a continuous theoretical model shows
important finite size effects. However, our data overlap fairly well with a discrete
model in the non--interacting case. The finite size effects found in the
non--interacting case could serve to estimate these effects in interacting systems. We
have more deeply studied
the structure of phase space analyzing other quantities like $\langle i \rangle$ (half
the average Hamming distance from a given configuration to the ground state), $S_N$ (the
``site occupation entropy'' see Eq.\ (\ref{S2})) and the valleys in configuration
space. For a given energy, $\langle i \rangle$ strongly depends on the type of
interactions while, for the systems considered, it does not depend on the system size. 
The ``site occupation entropy'', $S_N$, increases with the strength of
interactions and, for a long range one, with the size of the system.

\acknowledgments

We thank Prof.\ M. Pollak for useful discussions.
We would like to acknowledge financial support from the
 Spanish DGES, 
project numbers PB96-1118 and PB96-1120, and  from the
Fundaci\'on S\'eneca, Regi\'on de Murcia. A.P.G acknowledges  CajaMurcia 
for a grant.

\end{document}